\documentstyle[prl,aps,multicol,epsfig]{revtex}
\begin{document}
\draft

\renewcommand{\narrowtext}{\begin{multicols}{2}}
\renewcommand{\widetext}{\end{multicols}}
\newcommand{\be}{\begin{equation}}
\newcommand{\ee}{\end{equation}}
\newcommand{\lsim}   {\mathrel{\mathop{\kern 0pt \rlap
  {\raise.2ex\hbox{$<$}}}
  \lower.9ex\hbox{\kern-.190em $\sim$}}}
\newcommand{\gsim}   {\mathrel{\mathop{\kern 0pt \rlap
  {\raise.2ex\hbox{$>$}}}
  \lower.9ex\hbox{\kern-.190em $\sim$}}}
\def\be{\begin{equation}}
\def\ee{\end{equation}}
\def\ba{\begin{eqnarray}}
\def\ea{\end{eqnarray}}

\def\d{{\rm d}}
\def\i{{\rm i}}
\def\e{{\rm e}}

\def\ap{\approx}
\def\F{{\cal F}}
\def\G{{\cal G}}

\def\PLZ{P_{\rm LSZ}}

\def\L{{\mathcal L}}
\def\t{\vartheta}
\def\tm{\vartheta_{\rm m}}
\def\Dm{\Delta_{\rm m}}
\def\phi{\varphi}
\def\epsilon{\varepsilon}

\def\inpi2{\in [0\!:\!\pi/2]}

\font\menge=bbold9 scaled \magstep1
\def\nota#1{\hbox{$#1\textfont1=\menge $}}
\def\R{\nota R}

\title{Non-adiabatic level crossing in (non-) resonant neutrino oscillations} 

\author{M. Kachelrie{\ss}$^1$ and R. Tom\`as$^2$}

\address{$^1$TH Division, CERN, 1211 Geneva 23, Switzerland}
\address{$^2$Instituto de Fisica Corpuscular -- C.S.I.C.
          -- Universitat de Val\`encia,  
          46071 Val\`encia, Spain}

\date{March 30, 2001 --- CERN-TH/2001-098, IFIC/01-17}

\maketitle

\begin{abstract}
We study neutrino oscillations and the level-crossing probability
$\PLZ=\exp(-\gamma_n\F_n\pi/2)$ in power-law like potential
profiles $A(r)\propto r^n$. After showing that the resonance point
coincides only for a linear profile with the point of maximal violation of
adiabaticity, we point out that the ``adiabaticity'' parameter $\gamma_n$ 
can be calculated at an arbitrary point if the correction function $\F_n$
is rescaled appropriately. We present a new representation for the
level-crossing probability, $\PLZ=\exp(-\kappa_n\G_n)$, which allows a
simple numerical evaluation of $\PLZ$ in both the resonant and 
non-resonant cases and where $\G_n$ contains the full dependence of
$\PLZ$ on the mixing angle $\t$. 
As an application we consider the case $n=-3$ important for
oscillations of supernova neutrinos. 
\end{abstract}


\narrowtext


{\em Introduction---}%
The analytical study of non-adiabatic neutrino oscillations has a long
history. Soon after the discovery of ``resonant'' neutrino
flavor conversions by Mikheev and Smirnov~\cite{Mi85}, the leading
non-adiabatic effects were calculated for a linear potential
profile in Ref.~\cite{cross}. They were obtained in the form of the
Landau--St\"uckelberg--Zener (LSZ) crossing probability~\cite{LSZ}
\be 
 \PLZ =  \exp \left( -\frac{\pi\gamma}{2} \right) \,.
\ee
The adiabaticity parameter $\gamma$ for a linear profile is
\be  \label{gamma0}
 \gamma =   \frac{\Delta S^2}{2EC \:|\d\ln A/\d r|_0} \,,
\ee
where $E$ is the energy, $\Delta = m_2^2-m_1^2 >0$, $S=\sin(2\t)$,
$C=\cos(2\t)$, $\t\inpi2$ is the (vacuum) mixing angle and  
$A=2EV=2\sqrt{2}G_F N_e E$ is the induced mass squared term for the 
electron neutrino. The parameter $\gamma$ has to be evaluated at
the so-called resonance point $r_0$, i.e. the point where the mixing
angle in matter is $\tm=\pi/4$. For a linear profile, adiabaticity is
maximally violated for $\tm=\pi/4$ and, therefore, the probability that a
neutrino jumps from one branch of the dispersion relation to the other
is indeed maximal at $r_0$.  

Later, Kuo and Pantaleone derived in Ref.~\cite{Ku89} the LSZ crossing
probability for an arbitrary power-law like profile, $A\propto r^n$. 
They found that also in this case the
dependence on the neutrino masses and energies can be factored out,
while the effect of a non-linear profile can be encoded into a correction
function $\F_n$,
\be \label{WKB0}
 \PLZ =  \exp \left( -\frac{\pi\gamma_n}{2} \:  \F_n(\t) \right) \,,
\ee
which only depends on $\t$ and $n$. The ``adiabaticity'' parameter
$\gamma_n$ has to be evaluated still at $r_0$ although, as we will
show, it does not coincide with the point of maximal violation of
adiabaticity (pmva) for $n\neq 1$. An unsatisfactory feature of
Eq.~(\ref{WKB0}) is its restricted range of applicability: 
the resonance condition $\tm=\pi/4$ has a solution only
if $\t<\pi/4$ for neutrinos or if $\t>\pi/4$ for anti-neutrinos,
respectively. Therefore, it is not possible to calculate analytically,  
e.g., the survival probability of supernova anti-neutrinos  
in the quasi-adiabatic regime assuming a normal mass hierarchy.

The purpose of this Letter is twofold. First, we discuss the
physical significance of the resonance point $r_0$.
We show that the product $\gamma_n\F_n$
can be evaluated at an arbitrary point and conclude that the
``resonance'' point $r_0$ has, for a general profile $n\neq 1$, no
particular physical meaning: neither does it 
describe the point of maximal violation of adiabaticity nor is it
necessary to calculate the ``adiabaticity'' parameter $\gamma_n$ at $r_0$.
Second, we propose a new representation for $\PLZ$ that is
valid for all $\t$ and allows an easy numerical evaluation.
As an application, we consider the case $n=-3$ which is important for
oscillations of supernova neutrinos.

{\em Resonance point versus point of maximal violation of adiabaticity---}%
We use as starting point for our discussion the evolution equation for
the medium states $\tilde\psi$ first given in Ref.~\cite{Mi87}, 
\be    \label{S}
 \frac{\d}{\d r}
 \left( \begin{array}{c} \tilde\psi_1 \\ \tilde\psi_2 \end{array}
 \right) = 
 \left( \begin{array}{cc} \i\Dm/(4E) & -\tm^\prime \\
                          \tm^\prime & -\i\Dm/(4E)  \end{array}
 \right)  
 \left( \begin{array}{c} \tilde\psi_1 \\ \tilde\psi_2 \end{array}
 \right) 
\ee
Here, 
\be
 \Dm = \sqrt{\left(A-\Delta C\right)^2 + (\Delta S)^2}
\ee
denotes the difference between the effective mass of the two neutrino
states in matter, $\tm$ is the mixing angle in matter with
\be
 \tan 2\tm = \frac{\Delta S}{\Delta C -A}
\ee
and $\tm^\prime=\d\tm/\d r$. 

The evolution of the neutrino state is adiabatic at a given $r$, 
if the diagonal terms are large compared to the off-diagonal ones, 
$|\Dm|\gg|4E\tm^\prime|$. Thus the point where 
adiabaticity is maximally violated is given by the minimum of
$\Dm/\tm^\prime$. Differentiating~\cite{Fr00} 
\be 
 \frac{\Dm}{\tm^\prime} = \frac{2\Delta^2 S^2}{\sin^32\tm}\:
                          \frac{1}{\d A/ \d r}
\ee
for the case of a power-law profile, $A(r)\propto r^n$, we find the
minimum at 
\ba  \label{eq_pmva}
 && \cot(2\tm-2\t) + 2\cot(2\tm) \nonumber\\
 &-& 
 \frac{1}{n} \, \left[ \cot(2\tm-2\t) - \cot(2\tm) \right] = 0 \,.
\ea

For $n=1$, the pmva is at $\tm=\pi/4$ for all
$\t$. Thus, in the region where the resonance point is
well-defined, the pmva and $r_0$ coincide. 
In the general case, $n\neq 1$, the pmva  however agrees with the
resonance point only for $\t=0$. 
Finally, we recover the result of Ref.~\cite{Fr00} for an exponential
profile in the limit $n\to\pm\infty$.

\begin{figure}
\epsfig{file=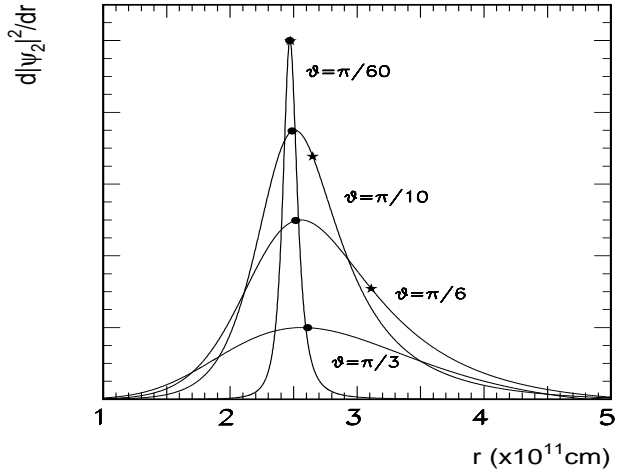,height=7.cm,width=9.cm,angle=0} 
\caption{\label{pmva2}
  Change of the survival probability $\d p(r)/\d r$
  of a neutrino produced at $r=0$ as $\nu_2$ together with the point
  of maximal violation of adiabaticity (dots)
   and the resonance point (stars) for
  a power-law profile $A\propto r^{-3}$. The height of the different
  curves is rescaled.}
\end{figure}

In Fig.~\ref{pmva2}, we show the change of the survival probability 
$\d p(r)/\d r=\d|\psi_2(r)|^2/\d r$ of a neutrino produced at $r=0$ as
$\nu_2$ as function of $r$ 
together with the point of maximal violation of adiabaticity predicted
by Eq.~(\ref{eq_pmva}) and the resonance point for a power-law profile
$A\propto r^{-3}$. 
It can be clearly seen that Eq.~(\ref{eq_pmva}) accurately describes 
the most probable position of the level crossing, while the resonance
point predicts a transition in less and less dense regions
until for $\t=\pi/4$ the concept of a resonant transition breaks down
completely. If the true profile is only approximately a power-law,
its exponent should be determined therefore by the region around the
pmva, not by the region around the resonance point.

{\em The correction functions $\F_n$---}%
We now recall briefly the calculation of the leading term to the
crossing probability within the WKB formalism~\cite{Ku89}. 
In the ultrarelativistic limit and
omitting an overall phase, the WKB formula results in
\be  \label{start}
 \ln\PLZ = -\frac{1}{E}\: \Im\int_{x_1(A_1)}^{x_2(A_2)} \d x \: \Dm  \,,     
\ee
where $A_2=\Delta e^{2\i\t}$ is the branch point of
$\Dm$ in the upper complex $x$ plane. 
We identify the physical coordinate $r\in [0:\infty]$ with the
positive part of the real $x$ axis, i.e. we consider a
neutrino state produced at small but positive $x$ propagating to
$x=\infty$. Then,   
a convenient choice for $A_1$ is to use the real part of $A_2$ for $C>0$, 
i.e. the ``resonance'' point $A_1=\Delta C$. However, we stress
that this choice has technical reasons and makes sense only for
$C>0$: consider for instance the simplest case $n=1$. Then both the
integration path chosen and the branch cut are for $C<0$ in the half-plane 
$\Re(x)<0$. The physical interpretation is therefore that an
anti-neutrino state created at small but negative $x$ 
propagates to $-\infty$. This case is however equivalent to a
neutrino state propagating with $C>0$ in the right half-plane and
therefore contains no new information. Thus, we expect the correction
functions $\F_n$ obtained with the integration path from $\Delta C$ to
$\Delta\e^{2\i\t}$ to be functions with period $\pi/4$ and to be valid
only in the resonant region.

After the substitution $A=\Delta(B+C)$, 
\be   \label{app}
 \ln\PLZ = -\frac{\Delta}{E}\: \Im\int_{0}^{\i S}
           \d B \: \frac{\d x}{\d B} \: (B^2 + S^2)^{1/2} \,,     
\ee
one has to expand the Jacobian $\d x/\d B$ in a power series
in order to solve the $B$ integrals.
Kuo and Pantaleone chose as expansion point the ``resonance'' point
$B_0=0$, because it leads to the simplest result. It is this choice, 
arbitrary from a physical point of view, that leads to the 
evaluation of the parameter $\gamma_n$ at $B_0=0$. 
In general, a change of $B_0$ in the definition of $\gamma_n$ will be
compensated by such a change in the correction function 
$\F_n$ that the physical observable $\PLZ$ is independent of
$B_0$~\cite{prep}.

The final result for the correction functions given by Kuo and
Pantaleone was

\be
 \F_n(\t, 0)  = 2 \sum_{m=0}^\infty 
 \left( \begin{array}{c} 1/n-1 \\ 2m \end{array} \right)
 \left( \begin{array}{c} 1/2 \\ m+1 \end{array} \right)
 \left( \tan(2\t) \right)^{2m} \!.
\ee
This series for $\F_n(\t,0)$ converges only for $\t<\pi/8$ and is
therefore not suited, even in the resonant region, for numerical
evaluation in the phenomenologically most interesting case of maximal
or nearly-maximal mixing. Representing the series as a hypergeometric
function, 
\be
 \F_n(\t, 0) = \,
 _2F_1\left( \frac{n-1}{2n}, \frac{2n-1}{2n};2;-\tan^2(2\t) \right) \,,
\ee
one can use, however, the Euler integral representation~\cite{F2} of $_2F_1$
as the analytical continuation for all $\t\in[0:\pi/4]$.

We now discuss the non-resonant case, $\t\in[\frac{\pi}{4}\!:\!\frac{\pi}{2}]$.
The expression $\gamma_n(0)\F_n(0,\t)$  becomes ill-defined in this
region for several reasons: 
the assumption $C\geq 0$ used in the derivation for
$\F_n(\t,0)$ is not fulfilled, and the ``resonance'' condition
$B=0$ or $A=\Delta C$ has no solution for $\t\in[\pi/4:\pi/2]$.
Moreover, the integration path used in Eq.~(\ref{app}) can be for $C<0$ in
the unphysical region $x<0$.
However, we know from the results above that the 
crossing probability is non-zero also in the non-resonant case. Also,
the WKB method should work independently of the sign of $C$ as long as
the evolution of the neutrino state is not strongly non-adiabatic.

We now show that it is not necessary to evaluate $\gamma_n\F_n$ at the 
resonance point $B_0=0$. Requiring the invariance of $\PLZ$ against
variations of $B_0$,   
\be
 \F_n\frac{\partial\gamma_n}{\partial B_0} + 
 \gamma_n\frac{\partial\F_n}{\partial B_0} = 0 \,,
\ee
we obtain with 
\be 
 \frac{\partial\gamma_n}{\partial B_0} = \frac{\gamma}{n(B_0+C)}
\ee 
a differential equation for $\d\F/\d B_0$. Its solution
\be
 \F_n(\t,B_0)= \left(\frac{C}{B_0+C}\right)^{1/n}\F_n(\t,0) 
\ee
allows us to rescale the correction functions obtained for $B_0=0$ to
arbitrary $B_0$. Therefore, we can calculate $\PLZ$ now also for the
non-resonant region in the two cases 
in which the function $\F_n$ is known for all $\t$.

\begin{figure}
\epsfig{file=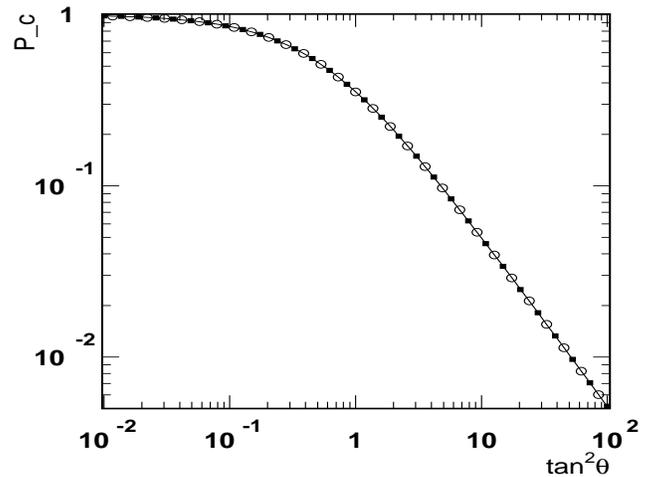,height=7.cm,width=9.cm,angle=0} 
\caption{\label{1/r}
Crossing probability $P_c$ for $A\propto 1/r$ with $R_0 V_0=0.2$:
numerically (solid line), with $\F_{-1}$ evaluated at $B_0=1$
(squares) and $\G_{-1}$ (circles).} 
\end{figure}

For a $1/r$ profile, $A(r)=2EV_0 (R_0/r)$, the correction function
is~\cite{Ku89} 
\be
\F_{-1}(0,\t)= \frac{\left(1-\tan^2(\t)\right)^2}{1+\tan^2(\t)} \,.
\ee
If we want to calculate the level-crossing probability in 
both the resonant and non-resonant cases, we have to choose
$B_0\geq 1$, e.g. 
\be
 \F_{-1}(1,\t)= \frac{1+C}{C} \F_{-1}(0,\t) \,;
\ee
the crossing probability follows as
\be \label{m1}
 \PLZ = \exp\left\{ -2\pi \, R_0 V_0 \, \sin^2(\t) \right\}  \,. 
\ee
As an example, we compare in Fig.~\ref{1/r} the results of a numerical
solution of the Schr\"odinger equation~(\ref{S}) with the analytical
calculation of the $\PLZ$ using the rescaled $F_{-1}$ function~\cite{f}. 
The agreement between the different methods is excellent.

In the limit $n\to\pm\infty$,
 which corresponds to an exponential potential profile, the 
scale factor $[C/(B_0+C)]^{1/n}$ goes to 1 and the correction
function becomes independent of $B_0$, as it should be. The resulting
crossing probability is
\be  \label{exp}
 \PLZ = \exp\left\{ -\frac{\pi\Delta R_0}{E} \: \sin^2(\t)  \right\} \,.
\ee
In Refs.~\cite{exp_early}, this expression was derived by solving 
Schr\"odinger's equation  directly. In these works, it was assumed that
the obtained expression is valid only in the resonant region and only
recently was it pointed out that it is valid for all
$\t$~\cite{exp_late}. Note also that $\ln\PLZ$ has the same, very
simple dependence on $\t$, for both $n=-1$ and $n=\pm\infty$.

{\em The correction functions $\G_n$---}%
We start directly from Eq.~(\ref{start}), but use now as integration
path in the complex $x$ plane the part of a circle centered at zero
starting at $A_1=\Delta$ and going to the end of the branch cut at
$A_2=\Delta\e^{2\i\t}$.  
Substituting $x=R_0(\Delta/A_0)^{1/n} \e^{\i\phi}$ in the case of a
potential $A=A_0 (r/R_0)^n$,
we can factor out the $\t$ dependence of $\PLZ$ into functions $\G_n$,   
\be
 \ln\PLZ = -\kappa_n \G_n(\t) \,,
\ee
where 
\be
 \kappa_n = 
                 \left( \frac{\Delta}{E} \right) \:
                 \left( \frac{\Delta}{A_0} \right)^{1/n} R_0
\ee
is independent of $\t$ and 
\be
 \G_n(\t) = \left| \; \Re
 \int_0^{2\t/n} \d\phi \:\e^{\i\phi}
 \left[\left(\e^{\i n\phi}-C\right)^2 + S^2 \right]^{1/2}  \right| \,.
\ee
The functions $\G_n$ are well suited for numerical evaluation and
always correspond to a neutrino state propagating in the physical part
of the $x$ plane, $x>0$. Therefore, they have, in contrast to the
$\F_n$ functions, the period $\pi/2$ and are valid for all $\t$.
In Fig.~\ref{1/r},  the results of this new representation are 
compared with those obtained above for the case $n=-1$.

{\em Oscillations of supernova neutrinos---}%
The potential profile $A(r)$ in supernova (SN) envelopes can be
approximated by a power law with $n\ap -3$, and
$V(r)= 1.5\times 10^{-9}$~eV $(10^9~{\rm cm}/r)^{-3}$~\cite{SN}.
Since only $\bar\nu_e$  were detected from SN~1987A and also in the
case of a future galactic SN the $\bar\nu_e$ flux will dominate the
observed neutrino signal, an analytical expression  for $P_c$ valid in the
non-resonant part of the mixing  space is especially useful~\cite{prep}. 
The probability for a $\bar\nu_e$ to arrive at the surface of the Earth
can be written as an incoherent sum of probabilities~\cite{fn},
\be
\label{Pee}
  P_{\bar e \bar e} = 
  \left(1-P_c\right)\cos^2\t + P_c\sin^2\t \,.
\ee
%
%

\begin{figure}
\epsfig{file=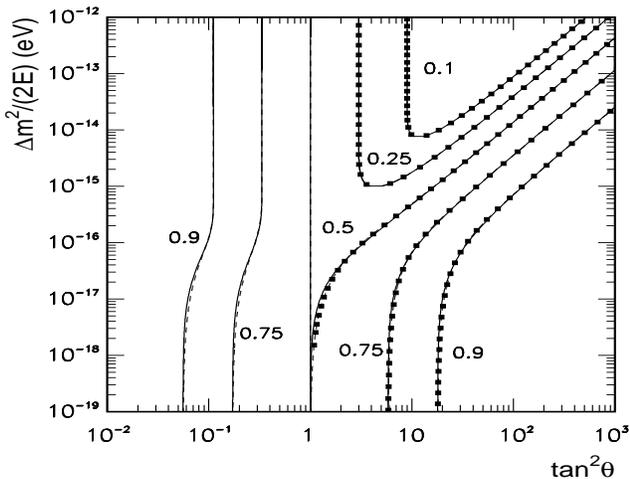,height=7.cm,width=9.cm,angle=0} 
\caption{\label{SN}
  Contours of constant survival probability $P_{\bar e\bar e}$,
  numerically (solid lines), with $\G_{-3}$ (dashed lines) and
  $\F_{-3}$ (squares, only for $\t>\pi/4$),  for $A\propto r^{-3}$ as
  given in the text.} 
\end{figure}
In Fig.~\ref{SN}, we compare the results of a numerical
solution of the Schr\"odinger equation~(\ref{S}) with the analytical
calculation of $P_{\bar e \bar e}$ using the $\G_{-3}$ and the
$\F_{-3}$ function. 
The latter is shown only for its range of applicability, $\t>\pi/4$.
The agreement between the two methods using the WKB approach is again  
(for $\t>\pi/4$) excellent. Generally, these two methods agree also very
well with the results of the numerical solution of the  Schr\"odinger
equation; there are only small deviations in the regions where the contours
change their slope.

{\em Summary---}%
We have discussed non-adiabatic neutrino oscillations in general
power-law potentials $A\propto r^n$. We have found that the
conventional splitting of $\ln\PLZ$ in an ``adiabaticity'' parameter
$\gamma_n$ evaluated at the resonance point and a correction function
$\F_n$ is misleading for $n\neq 1$: neither does the level-crossing
probability have a maximum at $r_0$ nor does this splitting allow the
calculation of $\PLZ$ in the non-resonant region. We have
proposed a new representation for $\PLZ$ that avoids these problems and
is hopefully useful for the investigations of oscillations of
supernova neutrinos.

{\em Acknowledgments---}%
We are grateful to Uli Nierste for helpful discussions. 
R.T. would like to thank the Theory Division at CERN for hospitality
and the Generalitat Valenciana for a grant.
This work was also supported by Spanish DGICYT grant PB98-0693, by the
European Commission TMR networks ERBFMRXCT960090 and
HPRN-CT-2000-00148, and by the European Science Foundation network N.~86.

\widetext


\begin{thebibliography}{00}
\bibitem{f}
We recall that the Landau--St\"uckelberg--Zener formula is 
only valid for small deviations from adiabaticity. Generally,
the WKB formula Eq.~(\ref{WKB0}) should be replaced~[4] by
$P_c = \frac{ \exp\left(-\frac{\pi\gamma}{2}\F\right) -
              \exp\left(-\frac{\pi\gamma}{2}\F^\prime\right)}
            { 1 - \exp\left(-\frac{\pi\gamma}{2}\F^\prime\right)}
$,
where $\F^\prime=\F/\sin^2\t$.
\bibitem{fn}
Note that the expressions
for anti-neutrinos can be obtained from those for neutrinos by
changing $\t\to\pi/2-\t$ in $P_c$.
\end{thebibliography}
\end{document}